\documentclass[12pt,letterpaper]{article}

\usepackage{amsmath}
\usepackage{amsfonts}
\usepackage{amssymb}
\usepackage{graphicx}
\def\be{\begin{equation}}
\def\ee{\end{equation}}

\begin{document}

\begin{flushright} {\footnotesize HUTP-05/A0038}\\ {\footnotesize MIT-CTP 3670}  \end{flushright}
\vspace{5mm}
\vspace{0.5cm}
\begin{center}

\def\thefootnote{\fnsymbol{footnote}}

{\Large \bf Limits on non-Gaussianities from WMAP data} \\[1cm]
{\large Paolo Creminelli$^{\rm a}$, Alberto Nicolis$^{\rm a}$, Leonardo Senatore$^{\rm b,d}$ \\
[.15cm]
Max Tegmark$^{\rm d}$   and Matias Zaldarriaga$^{\rm a,c}$}
\\[0.5cm]

{\small 
\textit{$^{\rm a}$ Jefferson Physical Laboratory, \\
Harvard University, Cambridge, MA 02138, USA}} 

\vspace{.2cm}

{\small 
\textit{$^{\rm b}$ Center for Theoretical Physics, \\
Massachusetts Institute of Technology, Cambridge, MA 02139, USA
}}

\vspace{.2cm}

{\small 
\textit{$^{\rm c}$ Center for Astrophysics, \\
Harvard University, Cambridge, MA 02138, USA
}}

\vspace{.2cm}

{\small 
\textit{$^{\rm d}$ Department of Physics, \\
Massachusetts Institute of Technology, Cambridge, MA 02139, USA
}}

\end{center}

\vspace{.8cm}

\hrule \vspace{0.3cm} 
{\small  \noindent \textbf{Abstract} \\[0.3cm]
\noindent
We develop a method to constrain the level of non-Gaussianity of density perturbations when the 3-point function is of the ``equilateral" type. Departures from Gaussianity of this form are produced by  single field models such as ghost or DBI inflation and in general by the presence of higher order derivative operators in the effective Lagrangian of the inflaton. We show that the induced shape of the 3-point function can be very well approximated by a factorizable form, making the analysis practical. We also show that, unless one has a full sky map with uniform noise,  in order to saturate the Cramer-Rao bound for the error on the amplitude of the 3-point function, the estimator must contain a piece that is linear in the data. We apply our technique to the WMAP data obtaining a constraint on the amplitude $f_{\rm NL}^{\rm equil.} $ of ``equilateral" non-Gaussianity: $-366 < f_{\rm NL}^{\rm equil.} < 238$ at $95\% \;{\rm C.L}$. We also apply our technique to constrain the so-called ``local" shape, which is predicted for example by the curvaton and variable decay width models. We show that the inclusion of the linear piece in the estimator improves the constraint over those obtained by the WMAP team, to $-27 < f_{\rm NL}^{\rm local} < 121$  at  $95\% \;{\rm C.L.}$

\vspace{0.5cm}  \hrule

\def\thefootnote{\arabic{footnote}}
\setcounter{footnote}{0}
\section{Introduction}
Despite major improvements in the quality of cosmological data over the last few years, there are very few observables that can characterize the early phases of the Universe, when density perturbations were generated. 
Deviations from a purely Gaussian statistics of density perturbations would be a very important constraint on models of early cosmology. In single field slow-roll inflation, the level of non-Gaussianity is sharply predicted 
\cite{Maldacena:2002vr,Acquaviva:2002ud} to be very small, less than $10^{-6}$. This is quite far from the present experimental sensitivity. On the other hand, many models have recently been proposed with a much higher level of non-Gaussianity, within reach of present or forthcoming data. For nearly Gaussian fluctuations, the quantity most 
sensitive to departures from perfect Gaussianity is the 3-point correlation function. 
In general, each model will give a different correlation 
between the Newtonian potential modes\footnote{Even with perfectly Gaussian primordial fluctuations, the observables, {\em e.g.} the temperature anisotropy, will not be perfectly Gaussian as a consequence of the non-linear relation between primordial perturbations and what we will eventually observe. These effects are usually of order $10^{-5}$ (see for example \cite{Creminelli:2004pv}) and thus beyond (but not much) present sensitivity. In the following we will disregard these contributions.}:
\be
\langle \Phi({\bf k}_1) \Phi({\bf k}_2) \Phi({\bf k}_3) \rangle = (2 \pi)^3 \delta^3
\big({\bf k}_1 + {\bf k}_2 + {\bf k}_3 \big)
F( k_1,  k_2 ,  k_3) \;.
\ee
The function $F$ describes the correlation as a function of the triangle shape in momentum space. 

The predictions for the function $F$ in different models divide quite sharply into two qualitatively different classes as a consequence of qualitatively different ways of producing correlations among modes \cite{Babich:2004gb}. The first possibility is that the source of density perturbations is not the inflaton but a second light scalar field
$\sigma$.
In this case non-Gaussianities are generated by the non-linear relation between the fluctuation $\delta\sigma$ of this field 
and the final perturbation $\Phi$ we observe. This non-linearity is {\em local} as it acts
when the modes are much outside the horizon; schematically we have  
$\Phi ({\bf x}) = A \, \delta\sigma({\bf x}) + B (\delta\sigma^2({\bf x}) - \langle\delta\sigma^2\rangle) + \ldots$ .
Even starting from a purely Gaussian $\delta\sigma$, the quadratic piece introduces a 3-point function 
for $\Phi$ of the form
\be
\label{eq:local}
F( k_1,  k_2 ,  k_3) = f_{\rm NL}^{\rm local} \cdot 2 \Delta_\Phi^2 \cdot \left(\frac1{k_1^3 k_2^3} + \frac1{k_1^3 k_3^3} + 
\frac1{k_2^3 k_3^3}\right) \;,
\ee
where $\Delta_\Phi$ is the power spectrum normalization,
$\langle \Phi({\bf k}_1) \Phi({\bf k}_2) \rangle = (2\pi)^3 \delta^3\big({\bf k}_1 + {\bf k}_2\big)
\Delta_\Phi \cdot k_1^{-3}$, which for the moment has been taken as exactly scale invariant, and where 
$f_{\rm NL}^{\rm local}$ is proportional to $B$.
Examples of this mechanism are the curvaton scenario \cite{Lyth:2002my} and the variable decay width model \cite{Zaldarriaga:2003my}, which naturally give rise to $f_{\rm NL}^{\rm local}$ greater than 10 and 5, respectively.

The second class of models are single field models with a non-minimal Lagrangian, where the correlation among modes is created by higher derivative operators \cite{Creminelli:2003iq,Arkani-Hamed:2003uz,Alishahiha:2004eh,Senatore:2004rj}. In this case, the correlation 
is strong among modes with comparable wavelength and it decays when we take one of $k$'s to zero
keeping the other two fixed. Although different models of this class give a different function $F$, all these functions are qualitatively very similar. We will call this kind of functions {\em equilateral}: as we will see, the signal is maximal for equilateral configurations in Fourier space, whereas for the local form (\ref{eq:local}) the most relevant configurations are the {\em squeezed} triangles with one side much smaller than the others.  

The strongest constraint on the level of non-Gaussianity comes from the WMAP experiment. The collaboration analyzed their data searching for non-Gaussianity of the {\em local} form (\ref{eq:local}), finding the data to be consistent with purely Gaussian statistics and placing limits on the parameter $f_{\rm NL}^{\rm local}$ \cite{Komatsu:2003fd}:
\be
\label{eq:fnlbound}
-58 < f_{\rm NL}^{\rm local} < 134 \quad {\rm at} \;95\% \;{\rm C.L.}
\ee

The main purpose of this paper is to perform a similar analysis searching for non-Gaussianities of the 
equilateral form. We can extend the definition of $f_{\rm NL}^{\rm local}$ in eq.~(\ref{eq:local}) to a 
generic function $F$ by setting the overall normalization on equilateral configurations:
\be
F(k,k,k) = f_{\rm NL} \cdot \frac{6 \Delta_\Phi^2}{k^6} \;.
\ee
In this way, two different models with the same $f_{\rm NL}$ will give the same 3-point function for equilateral configurations. For equilateral models in particular, the overall amplitude will be characterized by $f_{\rm NL}^{\rm equil.}$. Indirect constraints on $f_{\rm NL}^{\rm equil.}$ have been obtained  
starting from the limit of eq.~(\ref{eq:fnlbound}) in \cite{Babich:2004gb}, resulting in
 $|f_{\rm NL}^{\rm equil.}|\lesssim 800$;
 we will see that a dedicated analysis is, as expected, much more sensitive. Many of the equilateral models naturally predict sizeable values of the parameter $f_{\rm NL}^{\rm equil.}$: ghost inflation and DBI inflation  \cite{Arkani-Hamed:2003uz,Alishahiha:2004eh} tend to
have $f_{\rm NL}^{\rm equil.}\sim 100$, tilted ghost inflation  $f_{\rm NL}^{\rm equil.}\gtrsim 1$ \cite{Senatore:2004rj}, while
slow roll inflation with higher derivative coupling typically give $f_{\rm NL}^{\rm equil.}\lesssim 1$ \cite{Creminelli:2003iq}.

In Section \ref{factor} we discuss the main idea of the analysis. We will see that an optimal analysis
is numerically very challenging for a generic form of the function $F$, but simplifies dramatically 
if the function $F$ is factorizable in a sense that will be defined below. As all the equilateral forms
predicted in different models are (qualitatively and quantitatively) very similar, our approach will be 
to choose a factorizable function that well approximates this class. 
The analysis is further complicated by the breaking of rotational invariance:  a portion of the sky is fully masked by the presence of the galaxy 
and moreover each point is observed a different number of times. In Section \ref{optimal} we look for an 
optimal estimator for $f_{\rm NL}$. We discover that, unlike the rotationally invariant case, the minimum
variance estimator contains not only terms which are cubic in the temperature fluctuations, but also linear pieces.
These techniques are used to analyze the WMAP data in Section \ref{WMAP}. 
The result is that the data are compatible with Gaussian statistics. The limit for the equilateral models is 
\be
-366 < f_{\rm NL}^{\rm equil.} < 238  \quad {\rm at} \;95\% \;{\rm C.L.}
\ee
We also obtain a limit on $f_{\rm NL}^{\rm local}$
\be
-27 < f_{\rm NL}^{\rm local} < 121  \quad {\rm at} \;95\% \;{\rm C.L.}
\ee
which is a slight improvement with respect to (\ref{eq:fnlbound}) as a consequence of the above-mentioned additional 
linear piece in the estimator.

\section{\label{factor}A factorizable equilateral shape}
It is in principle straightforward to generalize the analysis for non-Gaussianities from the 
local model to another one. We start assuming that the experimental noise on the temperature maps is isotropic
and that the entire sky is observed (no Galactic and bright source mask); we will relax these assumptions in 
the next Section. In this case it can be proved that an optimal estimator ${\cal{E}}$ for 
$f_{\rm NL}$ exists \cite{Heavens:1998jb,Babich:2005en}, that is an estimator which saturates the Cramer-Rao inequality and thus 
gives the strongest possible constraint on the amplitude of the non-Gaussian signal. 
The estimator ${\cal{E}}$ is a sum of terms cubic in the temperature fluctuations, each term weighted by 
its signal to noise ratio:
\be
{\cal{E}} = \frac1N \cdot \sum_{l_i m_i} \frac{\langle a_{l_1 m_1}a_{l_2 m_2}a_{l_3 m_3} 
\rangle}{C_{l_1}C_{l_2}C_{l_3}} \; a_{l_1 m_1}a_{l_2 m_2}a_{l_3 m_3} \;.
\ee
Given the assumptions, the power spectrum $C_l$ (which is the sum of the CMB signal and noise) 
is diagonal in Fourier space; $N$ is a normalization factor which makes the estimator unbiased.
From rotational invariance we can simplify the expression introducing the Wigner $3j$ symbols to
\be
\label{eq:estfull}
{\cal{E}} = \frac1N \cdot \sum_{l_i m_i} \left(\begin{array}{ccc} l_1 & l_2 & l_3 
\\ m_1 & m_2 & m_3 \end{array}\right) \frac{B_{l_1 l_2 l_3}}{C_{l_1}C_{l_2}C_{l_3}} \;
a_{l_1 m_1}a_{l_2 m_2}a_{l_3 m_3}\;,
\ee
where $B_{l_1 l_2 l_3}$ is the angle-averaged bispectrum which contains all the information about the
model of non-Gaussianity we are considering. If $B_{l_1 l_2 l_3}$ is calculated for $f_{\rm NL} = 1$,
then the normalization factor $N$ is given by
\be
\label{eq:N}
N = \sum_{l_1 l_2 l_3} \frac{(B_{l_1 l_2 l_3})^2}{C_{l_1}C_{l_2}C_{l_3}} \;.
\ee
We now have to relate the angle-averaged bispectrum to the underlying correlation among 3d modes $F(k_1,k_2,k_3)$.
After some manipulations following \cite{Wang:1999vf}, the estimator takes the form 
\begin{eqnarray}
\label{eq:explestim}
{\cal{E}} & = & \frac1N \cdot \sum_{l_i m_i} \int 
d^2\hat{n} \; Y_{l_1 m_1}(\hat{n}) Y_{l_2 m_2}(\hat{n})Y_{l_3 m_3}(\hat{n}) 
\int \limits^{\infty}_0 r^2 dr \; j_{l_1}(k_1r) j_{l_2}(k_2r) j_{l_3}(k_3r) \; C_{l_1}^{-1} C_{l_2}^{-1} C_{l_3}^{-1} 
\nonumber \\ & & \int \frac{2 k^2_1 dk_1}{\pi} \frac{2 k^2_2 dk_2}{\pi} \frac{2 k^2_3 dk_3}{\pi} 
F(k_1,k_2,k_3) \Delta^T_{l_1}(k_1)
\Delta^T_{l_2}(k_2) \Delta^T_{l_3}(k_3) \; a_{l_1 m_1}a_{l_2 m_2}a_{l_3 m_3} \;,
\end{eqnarray}
where $\Delta^T_{l}(k)$ is the CMB transfer function which relates the $a_{lm}$ to the 
Newtonian potential $\Phi(k)$:
\be
a_{lm}=4\pi i^l \int \frac{d^3k}{(2\pi)^3} \Delta^T_l(k)  \Phi(k) Y^*_{lm}(\hat k) \;.
\ee
Equation (\ref{eq:explestim}) is valid for any shape of the 3-point function $F$. Unfortunately this expression
is computationally very challenging. The sums over $m$ can be taken inside the integrals and factorized, 
but we are still left with a triple sum over $l$ of an integral over the sphere. The calculation time 
grows as $N_{\rm pixels}^{5/2}$,  where the number of pixels $N_{\rm pixels}$ is of order $3\times 10^6$ for WMAP. 
This approach is therefore numerically too demanding. 

As noted in \cite{Wang:1999vf}, a crucial simplification is possible if the function $F$ is factorizable as a
product of functions of $k_1$, $k_2$ and $k_3$ or can be written as a sum of a small number of terms with this property.
In this case the second line of (\ref{eq:explestim}) becomes factored as the product of functions of each $l$ 
separately, so that now also the sum over $l$ can be done before integrating over the sphere. For example
if we assume that $F(k_1,k_2,k_3)= f_1(k_1) f_2(k_2) f_3(k_3)$, the estimator simplifies to 
\be
\label{eq:simple}
{\cal{E}} = \frac1N \cdot \int 
d^2\hat{n} \;
\int \limits^{\infty}_0 r^2 dr \; \prod^3_{i=1}\sum_{l_i m_i}\int \frac{2 k^2 dk}{\pi} j_{l_i}(k r)
f_i(k) \Delta^T_{l_i}(k) C_{l_i}^{-1} a_{l_i m_i} Y_{l_i m_i}(\hat{n})\;.
\ee
The calculation is obviously much faster now: it now scales like $N_{\rm pixels}^{3/2}$ and it is dominated by going back and forth between real and spherical harmonics space. From expression (\ref{eq:local}) we see that  this simplification is possible for the local shape, and it was indeed used for the analysis of the WMAP data in \cite{Komatsu:2003fd,Komatsu:2003iq}. 

Unfortunately, none of the ``equilateral models" discussed in the Introduction predicts a function $F$ which is factorizable (see some explicit expressions in \cite{Babich:2004gb}), so that it is not easy to perform an optimal analysis for a particular given model. However, all these models
give 3-point functions which are quite similar, so that it is a very good approximation to take a
factorizable shape function $F$ which is close to the class of functions we are interested in and perform
the analysis for this shape. In the limit $k_1 \to 0$ with $k_2$ and $k_3$ fixed, all the equilateral 
functions diverge as  $k_1^{-1}$ \cite{Babich:2004gb} (while the local form eq.~(\ref{eq:local}) goes 
as $k_1^{-3}$).
An example of a function which has this behavior, is symmetric in $k_1$, $k_2$ and $k_3$, and is a sum of 
factorizable functions (and is homogeneous of order $k^{-6}$, see below) is 
\footnote{Eq.~(\ref{eq:ours}) can be derived as follows. In order to make the divergence of $F$ mild
in the squeezed limit we can use at the numerator a quantity which goes to zero in the same limit.
The area of the triangle does the job, going like $k_1$ for $k_1 \to 0$.
The area can be expressed purely in terms of the sides through Heron's formula \cite{heron},
$A = \sqrt{s(s-k_1)(s-k_2)(s-k_3)}$, where $s=\frac12(k_1+k_2+k_3)$ is the semiperimeter. 
The first $s$ in the square root is irrelevant for our purposes, since it goes to a constant in the squeezed limit; 
we will therefore omit it. Also we want a sum of factorizable functions, so we get rid of the square root 
by considering $A^2$.
In conclusion, a function with all the features stated above is
\be
F(k_1,k_2,k_3) \propto \frac{(s-k_1)(s-k_2)(s-k_3)}{k_1^3 \, k_2^3 \, k_3^3} \; ,
\ee
which, once expanded, reduces exactly to eq.~(\ref{eq:ours}).
}
\be
\label{eq:ours}
F(k_1,k_2,k_3) = f_{\rm NL}^{\rm equil.} \cdot 6  
 \Delta_\Phi^2 \cdot \left(-\frac1{k_1^3 k_2^3} - 
\frac1{k_1^3 k_3^3} - \frac1{k_2^3 k_3^3}  - \frac2{k_1^2 k_2^2 k_3^2} + \frac1{k_1 k_2^2 k_3^3}
+ (5 \; perm.) \right) \;,
\ee
where the permutations act only on the last term in parentheses. The mild divergence for $k_1 \to 0$ is 
achieved through a cancellation among the various terms. 

In figure \ref{fig:shapes}, we compare this function with the local shape. The dependence of both functions under a common rescaling of all $k$'s is 
fixed to be $\propto k^{-6}$ by scale invariance, so that we can factor out $k_1^{-6}$ for example.
Everything will now depend only on the ratios $k_2/k_1$ and $k_3/k_1$, which fix the shape of the 
triangle in momentum space. For each shape we plot $F(1,k_2/k_1,k_3/k_1) (k_2/k_1)^2 (k_3/k_1)^2$; 
this is the relevant quantity if we are interested in the relative importance of different triangular 
shapes. The square of this function in fact gives the signal to noise contribution of a particular 
shape in momentum space \cite{Babich:2004gb}. We see that for the function (\ref{eq:ours}), the signal 
to noise is concentrated on equilateral configurations, while squeezed triangles with one side much 
smaller than the others are the most relevant for the local shape.

\begin{figure}[th!!]             
\begin{center}
\includegraphics[width=12.5cm]{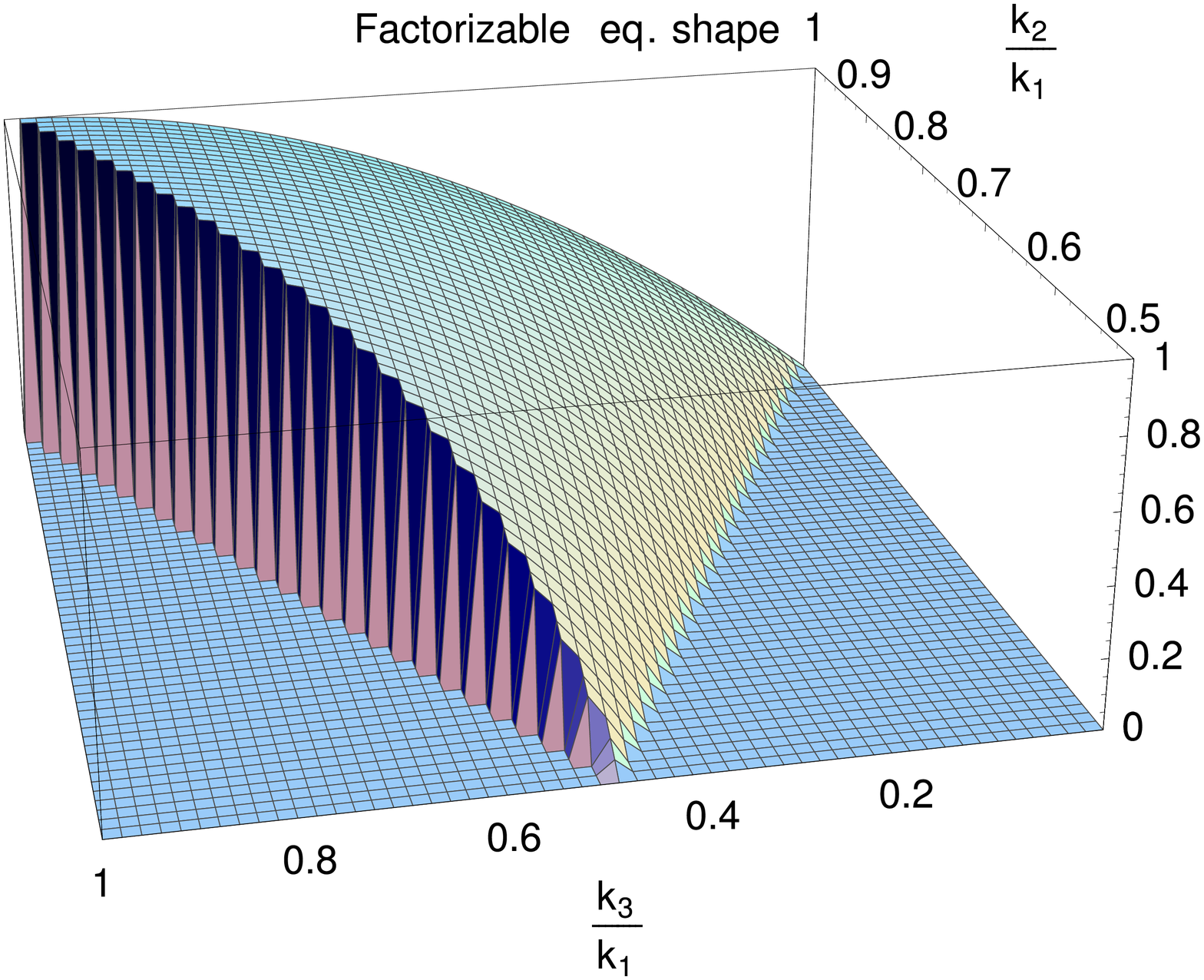}

\vspace{.7cm}
\includegraphics[width=12.5cm]{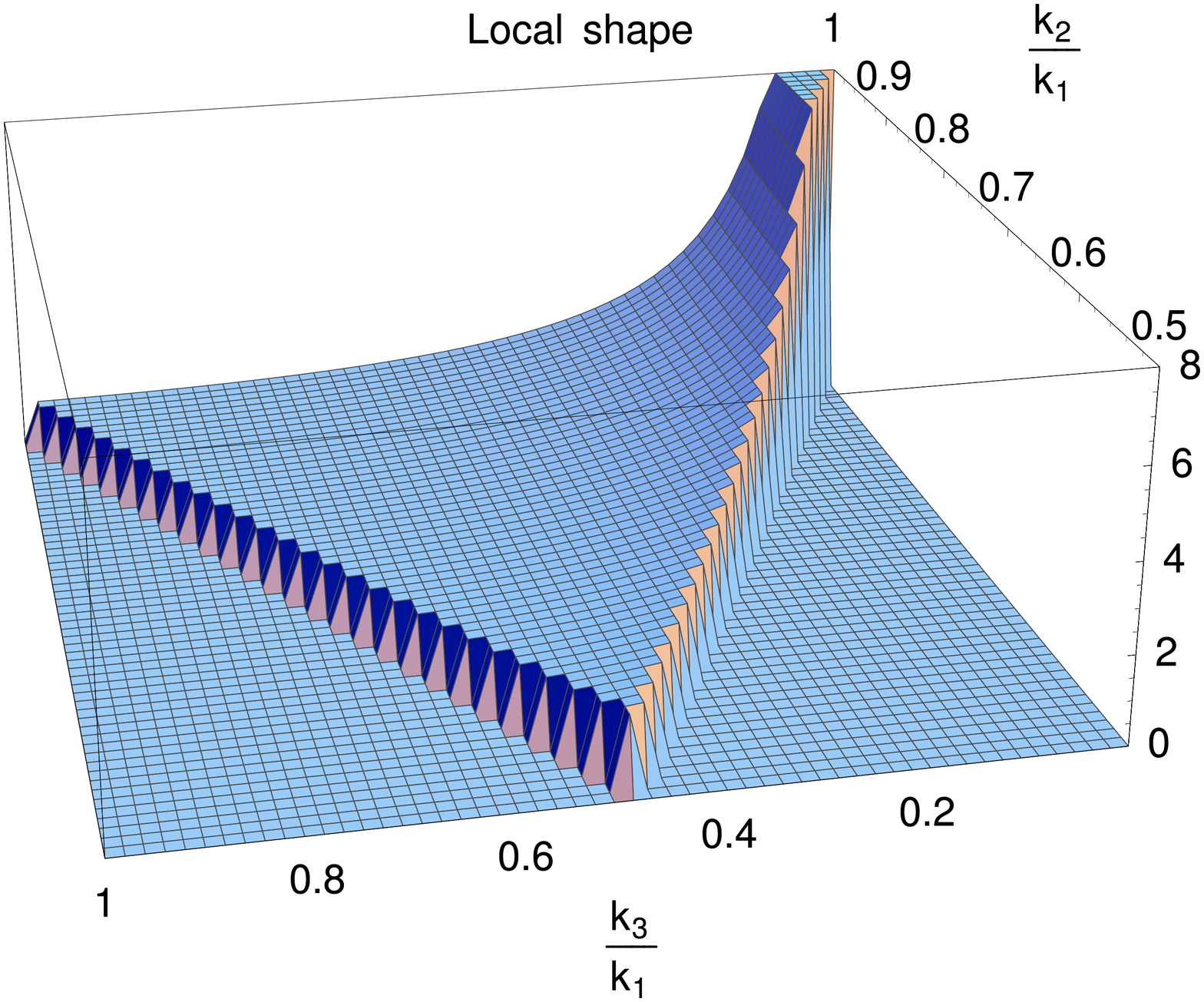}

\caption{\label{fig:shapes} \small  Plot of the function 
$F(1,\, k_2/k_1, \, k_3/k_1) (k_2/k_1)^2 (k_3/k_1)^2$ for the 
equilateral shape used in the analysis (top) and for the local shape (bottom). The functions are both 
normalized to unity for equilateral configurations $\frac{k_2}{k_1}= \frac{k_3}{k_1}=1$.
Since $F(k_1,k_2,k_3)$ is symmetric in its three arguments, 
it is sufficient to specify it for $k_1\ge k_2\ge k_3$, so $\frac{k_3}{k_1} \le \frac{k_2}{k_1} \le 1$ above.
Moreover, the triangle inequality says that no side can be longer than the sum of the other two,
so we only plot $F$ in the triangular region $1-\frac{k_2}{k_1}\leq \frac{k_3}{k_1} \leq\frac{k_2}{k_1} \le 1$ above, 
setting it to zero elsewhere.}
\end{center}
\end{figure} 

In figure \ref{fig:hddiff} we study the equilateral function predicted both in the presence of higher-derivative terms \cite{Creminelli:2003iq} and in DBI inflation \cite{Alishahiha:2004eh}. In the second part of the figure we show the difference between this function and the factorizable one used in our analysis. 
We see that the relative difference is quite small. The same remains true for other equilateral shapes
(see \cite{Babich:2004gb} for the analogous plots for other models).
\begin{figure}[t!]             
\begin{center}
\includegraphics[width=12.5cm]{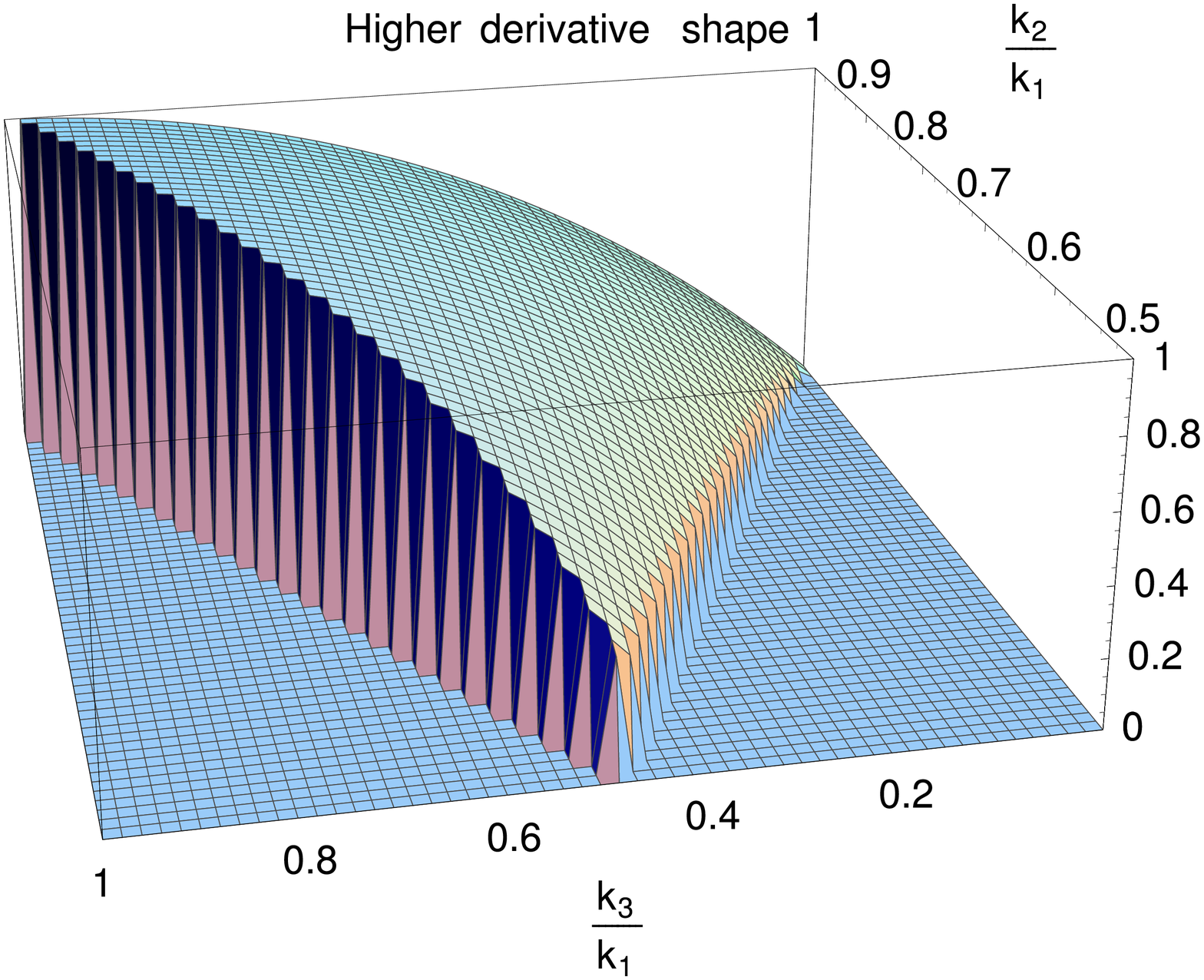}
\vspace{.7cm}

\includegraphics[width=12.5cm]{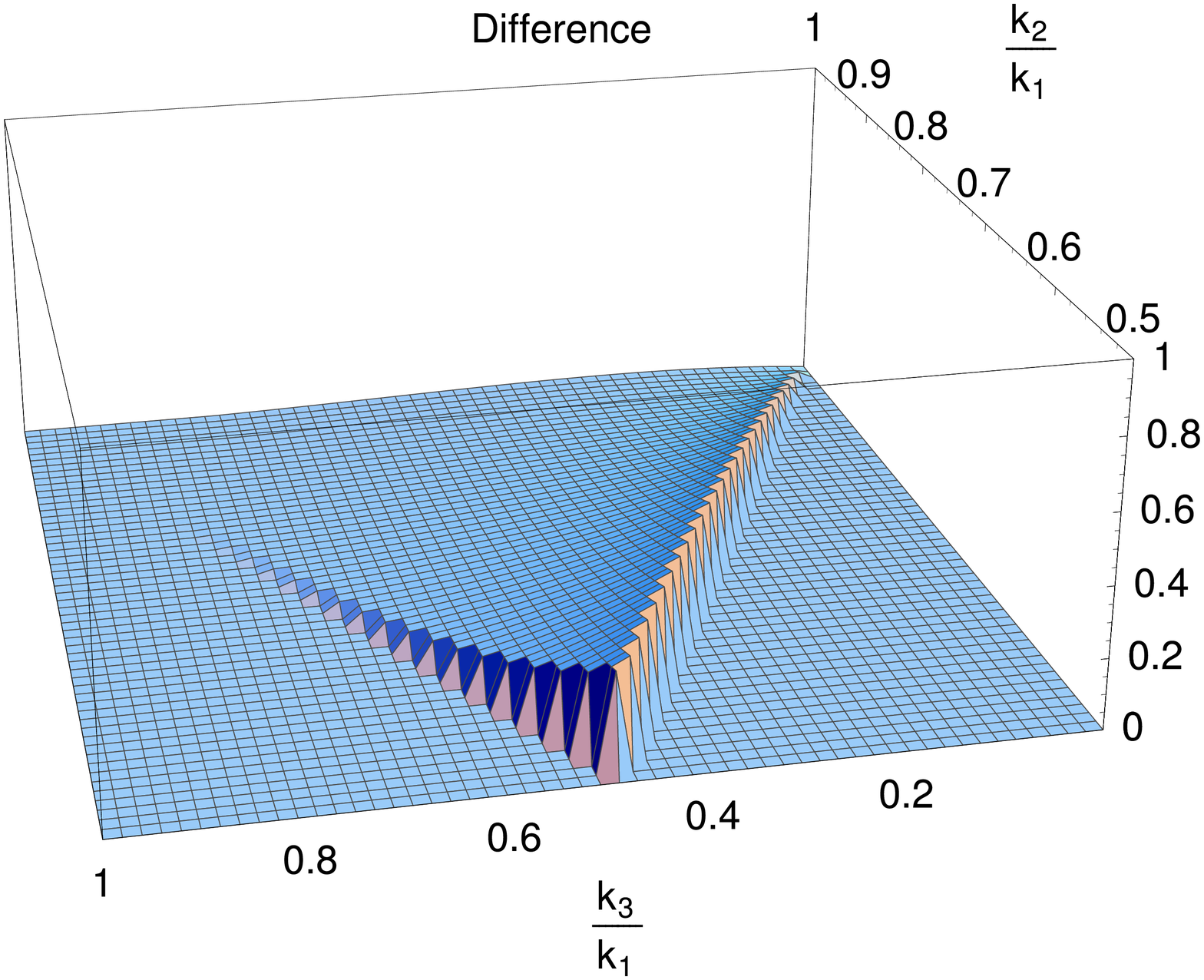}
\caption{\label{fig:hddiff} \small  Top. Plot of the function $F(1,\, k_2/k_1, \, k_3/k_1) (k_2/k_1)^2 (k_3/k_1)^2$
predicted by the higher-derivative \cite{Creminelli:2003iq} and the DBI models \cite{Alishahiha:2004eh}. 
Bottom. Difference between the above plot and the analogous one (top of fig.~\ref{fig:shapes}) for the factorizable 
equilateral shape used in the analysis.}
\vspace{2cm}

\end{center}
\end{figure}

 In \cite{Babich:2004gb}, a ``cosine" between different shapes was defined which quantifies how different is the signal given by two distributions. The cosine is calculated from the scalar product of 
the functions in fig.~\ref{fig:shapes}. We can think about this cosine as a sort of correlation coefficient: if the cosine is close to $1$ the two shapes are very difficult to distinguish experimentally
and an optimal analysis for one of them is very good also for the other. On the other hand, a small cosine means 
that, once non-Gaussianities are detected, there is a good chance to distinguish the two functions and 
that an optimal analysis for one shape is not very good for the other.  The cosine between our template shape 
and the functions predicted by equilateral models is very close to one ($0.98$ with the ghost inflation
\cite{Arkani-Hamed:2003uz} prediction and $0.99$ for higher derivative/DBI models \cite{Creminelli:2003iq, Alishahiha:2004eh}). This means that the error introduced in the analysis by the use of the factorizable shape instead of the correct prediction for a given equilateral model is at the percent level. On the other hand, as evident from figure \ref{fig:shapes}, our 
template shape is quite different from the local model --- the cosine is merely $0.41$.

All these numbers are obtained in 3 dimensions assuming that we can directly measure the fluctuations
with a 3d experiment like a galaxy survey. The CMB anisotropies are a complicated projection from
the 3d modes. This makes it more difficult to distinguish different shapes, although  
the picture remains qualitatively the same.

Let us proceed with the study of our template shape.
To further simplify the estimator in eq.~(\ref{eq:explestim}), we define the functions
\begin{eqnarray}
\alpha_l(r) & \equiv & \frac2\pi \int_0^{+ \infty} \!\!\! dk \; k^2 \, \Delta^T_l(k) j_l(k r) \\
\beta_l(r) & \equiv & \frac2\pi \int_0^{+ \infty} \!\!\! dk \; k^{-1} \, \Delta^T_l(k) j_l(k r)\Delta_\Phi \\ 
\gamma_l(r) & \equiv & \frac2\pi \int_0^{+ \infty} \!\!\! dk \; k \, \Delta^T_l(k) j_l(k r) \Delta_\Phi^{1/3}\\
\delta_l(r) & \equiv & \frac2\pi \int_0^{+ \infty} \!\!\! dk  \, \Delta^T_l(k) j_l(k r)\Delta_\Phi^{2/3} \;.
\end{eqnarray}
 We use this strange ordering of the functions to keep the notation compatible with \cite{Komatsu:2003fd,Komatsu:2003iq}, where the functions $\alpha$ and $\beta$ were
 introduced for the analysis of the local shape. To evaluate ${\cal{E}}$, we start from the spherical harmonic coefficients of the map $a_{lm}$ and calculate
 the four maps 
 \begin{eqnarray}
A(r, \hat n)  \equiv  \sum_{l m} \frac{\alpha_l(r)}{C_l} Y_{lm}(\hat n) a_{lm} \;, & & B(r, \hat n)  \equiv  \sum_{l m} \frac{\beta_l(r)}{C_l} Y_{lm}(\hat n) a_{lm}\;, \\  C(r, \hat n)  \equiv  
\sum_{l m} \frac{\gamma_l(r)}{C_l} Y_{lm}(\hat n) a_{lm} \;, & & D(r, \hat n)  \equiv  \sum_{l m} \frac{\delta_l(r)}{C_l} Y_{lm}(\hat n) a_{lm} \;.
\end{eqnarray} 
Now the estimator ${\cal{E}}$ is given by
\be
{\cal{E}} = -\frac{18}N \int  r^2 dr \int d^2 \hat n \; \left[ A(r, \hat n) B(r, \hat n)^2 + \frac23 
D(r, \hat n)^3 - 2 B(r, \hat n)C(r, \hat n)D(r, \hat n) \right] 
\ee
and the normalization $N$ can be calculated from (\ref{eq:N}) using the explicit form
\begin{eqnarray}
&& \!\!\!\!\!\!\!\!B_{l_1 l_2 l_3}  =  \sqrt{\frac{(2 l_1+1) (2 l_2+1)(2 l_3+1)}{4 \pi}} \left(\begin{array}{ccc} l_1 & l_2 & l_3 
\\ 0 & 0 & 0 \end{array}\right)  \times \\ & & 6 \int_0^\infty \!\!r^2 dr \left[ -\alpha_{l_1}(r) \beta_{l_2}(r)\beta_{l_1}(r) 
+ (2 \;{\it perm.}) - 2 \delta_{l_1}(r)\delta_{l_2}(r) \delta_{l_3}(r) + \beta_{l_1}(r)\gamma_{l_2}(r)\delta_{l_3}(r) + ê  (5 \;{\it perm.})\right].  \nonumber
\end{eqnarray}
The functions $\alpha$, $\beta$, $\gamma$ and $\delta$ used in the analysis can be obtained numerically  starting 
from the transfer functions $\Delta^T_{l}(k)$, which can be computed given a particular cosmological
model with publicly available software as CMBFAST  \cite{Seljak:1996is}. Some plots of these functions 
are given in fig.~\ref{fig:abcd}, where we choose values of $r$ close to $\tau_0-\tau_R$ (conformal time difference between recombination and the present), as these give the largest contribution to the estimator. 
The oscillatory behavior induced by the transfer 
function $\Delta^T_l(k)$ is evident, with a peak at $l\sim 200$. The factors of $l(l+1)$ and $(2l+1)$ in the $\beta$ and 
$\delta$ functions, respectively, can be understood
from the behavior at low $l$'s, in the Sachs-Wolfe regime. Here the transfer function can be approximated by $\Delta^T_l(k)=-j_l\left(k (\tau_0-\tau_R)\right)/3$, and we then see that 
$\beta_l(\tau_0-\tau_R)\propto\int^\infty_0 dk\; k^{-1}\; j_l^2(k (\tau_0-\tau_R)) \propto 1/\left(l(l+1)\right)$ and 
$\delta_l(\tau_0-\tau_R)\propto\int^\infty_0 dk\; j_l^2(k (\tau_0-\tau_R)) \propto 1/(2l+1)$. From these expressions we also see that the function $\beta$ (the only one which is dimensionless) is very similar to the $C^{{\rm cmb}}_l$'s as, in the Sachs-Wolfe regime,
$C^{{\rm cmb}}_l\simeq \frac{2}{9\pi}\int^\infty_0 dk\; k^{-1}\; \Delta_\Phi\; j^2_l(k (\tau_0-\tau_R))$.

\begin{figure}[t!]             
\begin{flushleft}
\includegraphics[width=8.2cm]{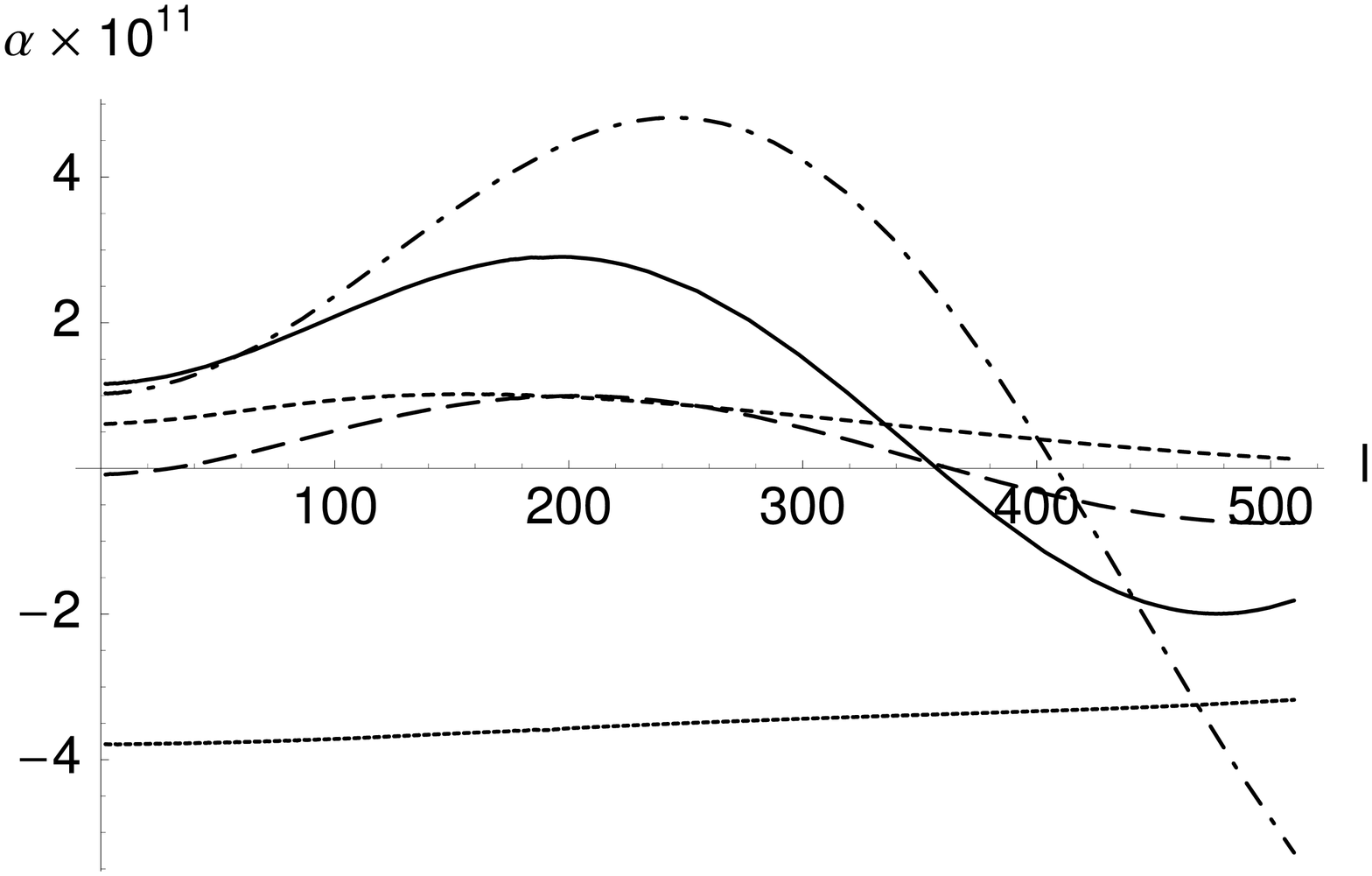}
\includegraphics[width=8.2cm]{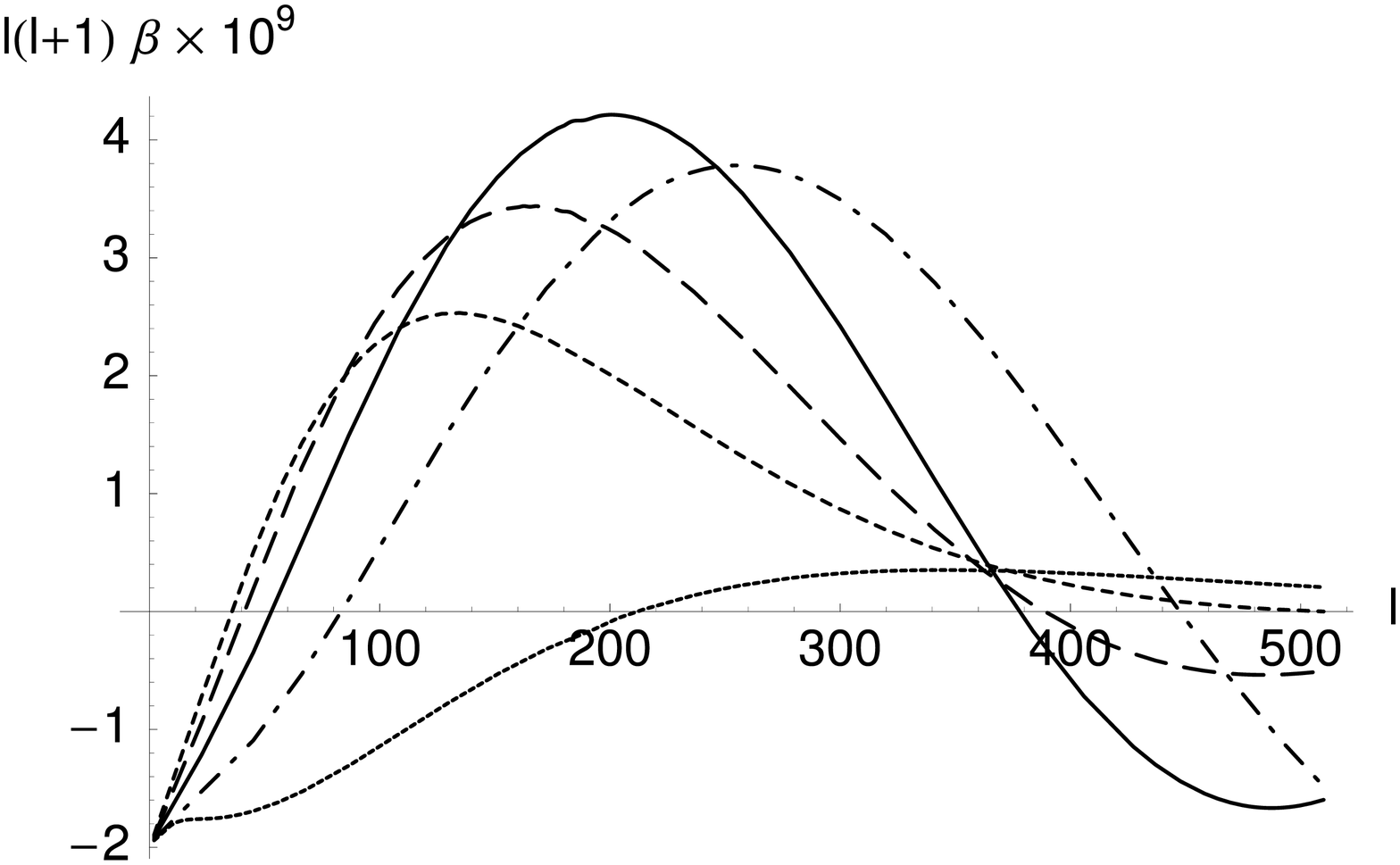}

\vspace{1cm}
\includegraphics[width=8.2cm]{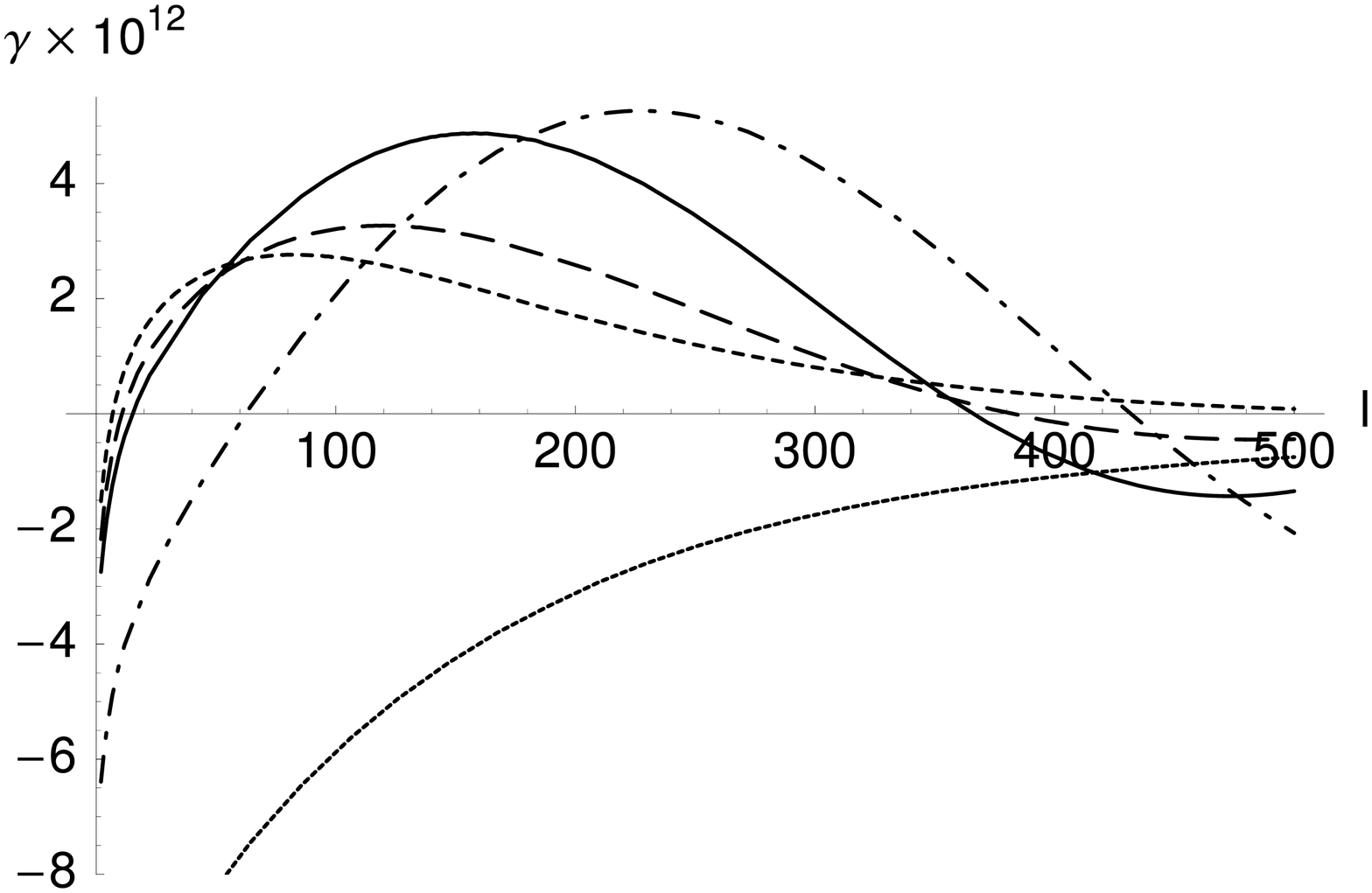}
\includegraphics[width=8.2cm]{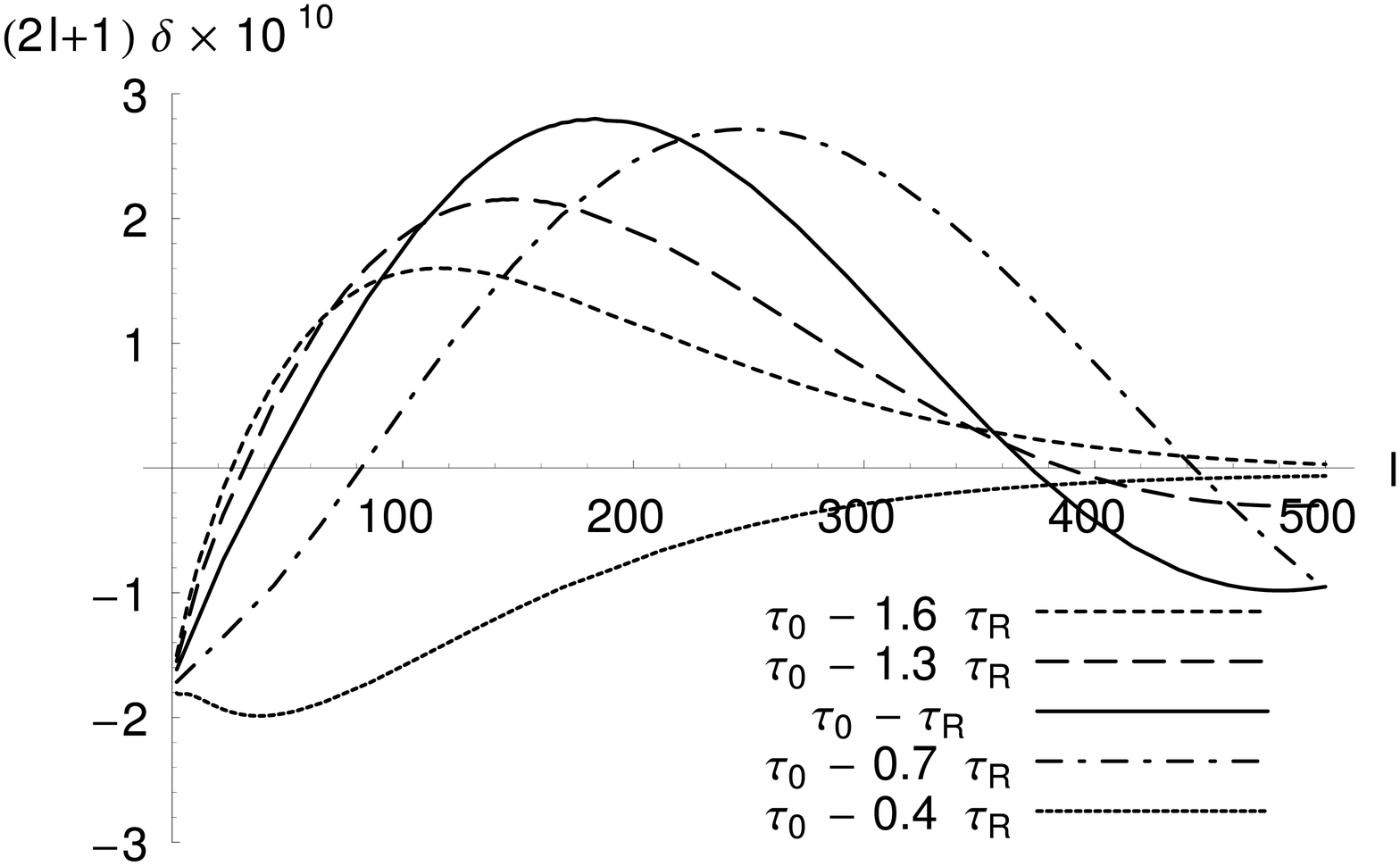}

\caption{\label{fig:abcd} \small The functions $\alpha_l(r)$ (in units of ${\rm Mpc}^{-3}$), $\beta_l(r)$ (dimensionless), $\gamma_l(r)$ (in units of ${\rm Mpc}^{-2}$), and $\delta_l(r)$ (in units of ${\rm Mpc}^{-1}$) are shown for various radii $r$
as functions of the multipole number $l$. The cosmological parameters are the same ones used in the analysis:
$\Omega_b h^2 = 0.024$, $\Omega_m h^2 = 0.14$, $h =0.72$, $\tau = 0.17$. With these parameters, 
the present conformal time $\tau_0$ is $13.24$ Gpc and the recombination time $\tau_R$ is  $0.27$ Gpc.}
\end{flushleft}
\end{figure}

\section{\label{optimal}Optimal estimator for $f_{\rm NL}$}

There are two quite general experimental complications that make the estimator $\cal{E}$ defined in the last Section non-optimal. First, foreground emission from the Galactic plane and from isolated bright sources contaminates a 
substantial fraction of sky, which must 
be masked out before the analysis \cite{Bennett:2003ca}. This projection which can be accomplished
by giving very large noise to regions that are affected by foregrounds, besides reducing the available 
amount of data, has the important effect of breaking rotational invariance, so that the signal covariance 
matrix $C^{\rm cmb}_{l_1 m_1, l_2 m_2}$ of the masked sky becomes non-diagonal in multipole space.
Second, the noise level is not the same in different regions of the sky, because the experiment looks at 
different regions for different amounts of time. This implies that also the noise covariance 
matrix $C^{\rm noise}_{l_1 m_1, l_2 m_2}$ is non-diagonal in multipole space.
The importance of the anisotropy of the noise was pointed out in \cite{Komatsu:2003fd}, as it 
caused an increase of the variance of the estimator ${\cal{E}}$ at high $l$'s.

For these reasons, it is worth trying to generalize the estimator ${\cal{E}}$. We consider estimators for 
$f_{\rm NL}$ (for a given shape of non-Gaussianities) which contain
both trilinear and linear terms in the $a_{l m}$'s. In the rotationally invariant case, a linear piece 
in the estimator can only be proportional to the monopole, which is unobservable. However, in the absence of rotational invariance, linear terms can be relevant. In this class, it is straightforward to prove that the unbiased 
estimator with the smallest variance is given by
\begin{eqnarray}
\label{eq:estlin}
{\cal{E}}_{\rm lin}(a)& = &\frac{1}{N}\sum_{l_i m_i}\Big(\langle a_{l_1 m_1}a_{l_2 m_2}a_{l_3 m_3}\rangle_1 \;
C^{-1}_{l_1 m_1, l_4 m_4}C^{-1}_{l_2 m_2, l_5 m_5}C^{-1}_{l_3 m_3, l_6 m_6}
a_{l_4 m_4}a_{l_5 m_5}a_{l_6 m_6}  \\ \nonumber 
&& -3 \;\langle a_{l_1 m_1}a_{l_2 m_2}a_{l_3 m_3}\rangle_1\;C^{-1}_{l_1 m_1,l_2 m_2} C^{-1}_{l_3 m_3,l_4 m_4}
a_{l_4 m_4} \Big) \;,
\end{eqnarray}
where $N$ is a normalization factor:
\begin{equation}
N=\sum_{l_i m_i}\langle a_{l_1 m_1}a_{l_2 m_2}a_{l_3 m_3}\rangle_1 \;C^{-1}_{l_1 m_1, l_4 m_4}
C^{-1}_{l_2 m_2, l_5 m_5}C^{-1}_{l_3 m_3, l_6 m_6}\langle a_{l_4 m_4}a_{l_5 m_5}a_{l_6 m_6}\rangle_1 \;.
\end{equation}
The averages $\langle\ldots\rangle_1$ are taken with $f_{\rm NL} = 1$. 
As we stressed, the covariance matrix $C = C^{\rm cmb} + C^{\rm noise}$ contains the effects of the mask
projection and the anisotropic noise and it is thus non-diagonal in multipole space.

It turns out that the estimator ${\cal{E}}_{\rm lin}$ saturates the Cramer-Rao bound, {\em i.e.} it is not only 
optimal in its class, but it is also the minimum variance unbiased estimator among all the possible ones.
To prove this, we expand the probability distribution in the 
limit of weak non-Gaussianity (which is surely a good approximation for the CMB) using the Edgeworth expansion 
\cite{Juszkiewicz:1993hm,Taylor:2000hq,Bernardeau:1994aq}.
\begin{equation}
\label{eq:PDF}
P(a|f_{\rm NL})=\Big(1- f_{\rm NL}\sum_{l_i m_i} \langle a_{l_1 m_1}a_{l_2 m_2}a_{l_3 m_3}\rangle_1
\frac{\partial}{\partial a_{l_1 m_1}}\frac{\partial}{\partial a_{l_2 m_2}}
\frac{\partial}{\partial a_{l_3 m_3}}
\Big)\frac{e^{-\frac12{\sum a^*_{l_4m_4}C^{-1}_{l_4m_4,l_5 m_5}a_{l_5 m_5}}}}
{\sqrt{(2\pi)^N {\rm det}(C)}} \;.
\end{equation}
Now, following the same arguments of \cite{Babich:2005en} which considered the same problem
in the rotationally invariant case, the estimator is optimal (in the sense that its variance saturates
the Cramer-Rao bound), if and only if the following condition is satisfied:
\begin{equation}
\label{eq:optimal}
\frac{d \log P(a|f_{\rm NL})}{d \,f_{\rm NL}}=F(f_{\rm NL})\left({\cal{E_{\rm lin}}}(a)-f_{\rm NL} \right) \;,
\end{equation}
where $F(f_{\rm NL})$ is a generic function\footnote{When 
eq.~(\ref{eq:optimal}) is satisfied, the function $F$ turns out to be the Fisher 
information for the parameter $f_{\rm NL}$, $F(f_{\rm NL}) =\langle (d \log P(a|f_{\rm NL})/ d f_{\rm NL})^2  
\rangle$.} of the parameter $f_{\rm NL}$.
From the probability distribution (\ref{eq:PDF}) it is easy to check that the estimator ${\cal{E}_{\rm lin}}$ 
is proportional to $d \log P(a|f_{\rm NL})/d \,f_{\rm NL}$ in the limit of small $f_{\rm NL}$. We conclude 
that ${\cal{E_{\rm lin}}}$ is an optimal estimator for a nearly Gaussian distribution.

We now want to make some approximations to the optimal estimator ${\cal{E}_{\rm lin}}$ to make
it numerically easier to evaluate. The full inversion of the covariance matrix is computationally rather cumbersome 
(although doable as the matrix is with good approximation block diagonal \cite{Hinshaw:2003ex}). We therefore approximate 
$C^{-1} a$ in the trilinear term of eq.~(\ref{eq:estlin}) by masking out the sky before computing the $a_{lm}$'s and taking $C$ 
as diagonal: 
\be
{\rm trilinear} \to \frac1N\sum_{l_i m_i} \frac{\langle a_{l_1 m_1}a_{l_2 m_2}a_{l_3 m_3}\rangle_1} {C_{l_1}C_{l_2}
C_{l_3}} \; a_{l_1 m_1} a_{l_2 m_2} a_{l_3 m_3} \;,
\ee
where $\langle a_{l_1 m_1}a_{l_2 m_2}a_{l_3 m_3} \rangle_1$ is still given by 
the full sky expressions of the last Section.
Once we have made this approximation for the trilinear term
it is easy to prove that the choice for the linear term which minimizes the variance is
\be
{\rm linear} \to-\frac3N \sum_{l_i m_i} \frac{\langle a_{l_1 m_1}a_{l_2 m_2}a_{l_3 m_3}\rangle_1} {C_{l_1}C_{l_2}
C_{l_3}} \;C_{l_1 m_1,l_2 m_2}\, a_{l_3 m_3} \;.  
\ee
Note that we must not approximate the covariance matrix $C$ in the numerator as diagonal. 
This would leave only a term proportional to the monopole $a_{00}$ (which is unobservable), as in the rotationally 
invariant case. 

The normalization factor $N$ is  given by 
\be
N = f_{\rm sky} \sum_{l_1 l_2 l_3} \frac{(B_{l_1 l_2 l_3})^2}{C_{l_1}C_{l_2}C_{l_3}} \;,
\ee
where $f_{\rm sky}$ is the fraction of the sky actually observed. As shown in \cite{Komatsu:2002db}, 
this correctly takes into account the reduction of data introduced by the mask for multipoles much 
higher than the inverse angular scale of the mask. The accuracy of this approximation has been checked
on non-Gaussian simulations for the local shape in \cite{Komatsu:2003fd}.

Let us try to understand qualitatively the effect of the linear correction. Take a large region of the sky
that has been observed many times so that its noise level is low. This region will therefore have
a small-scale power lower than average. Now, in a given realization depending on how the large scale modes look like, this long-wavelength modulation 
of the small-scale power spectrum may be ``misinterpreted'' by the trilinear estimator 
as a non-Gaussian signal. Indeed, for the local shape, most of the signal comes precisely from the correlation 
between long wavelength modes and the small scale power 
\cite{Babich:2004gb}. On the other hand, for equilateral shapes, as we noted, the signal is quite low
on squeezed configurations so that we expect this effect to be small. The linear term measures the correlation 
between a given map and the anisotropies in the power spectrum, thus correcting for this spurious signal. Clearly the spurious correlation is zero on average, but the effect increases the variance of the estimator.

We will apply the linear correction of the estimator only for the local shape, since, as we will verify later, 
it gives a very small
effect for equilateral shapes. Following the same steps as in the last Section to factorize the trilinear
term in the estimator, we get an explicit expression for the linear piece in the local case: 
\be
-\frac3N \int  r^2 dr \int d^2\hat n \sum_{l_3 m_3} \left( 2 S_{AB}(\hat n,r) 
\frac{\beta_{l_3}(r)}{C_{l_3}}Y_{l_3,m_3}(\hat n)+S_{BB}(\hat n,r) 
\frac{\alpha_{l_3}(r)}{C_{l_3}}Y_{l_3,m_3}(\hat n)\right) a_{l_3 m_3} \;,
\ee
where the two maps $S_{AB}(\hat n,r)$ and $S_{BB}(\hat n,r)$ are defined as
\begin{equation}
\label{eq:lin1}
S_{AB}(\hat n,r)=\sum_{l_i m_i} \frac{\alpha_{l_1}(r)}{C_{l_1}}Y_{l_1,m_1}(\hat n) 
\frac{\beta_{l_2}(r)}{C_{l_2}}Y_{l_2,m_2}(\hat n) \langle a_{l_1 m_1} a_{l_2 m_2}\rangle\;,
\end{equation}
\begin{equation}
\label{eq:lin2}
S_{BB}(\hat n,r)=\sum_{l_i m_i} \frac{\beta_{l_1}(r)}{C_{l_1}}Y_{l_1,m_1}(\hat n) 
\frac{\beta_{l_2}(r)}{C_{l_2}}Y_{l_2,m_2}(\hat n) \langle a_{l_1 m_1} a_{l_2 m_2}\rangle \;.
\end{equation}
As discussed above, it is only the anisotropic part of the matrix $\langle a_{l_1 m_1} a_{l_2 m_2}\rangle$ that gives a
contribution to the linear part of the estimator. This matrix can be decomposed as
\be
\langle a_{l_1 m_1} a_{l_2 m_2}\rangle= 
\langle a^{{\rm cmb}}_{l_1 m_1} a^{{\rm cmb}}_{l_2 m_2}\rangle+\langle a^{{\rm noise}}_{l_1 m_1} 
a^{{\rm noise}}_{l_2 m_2}\rangle,
\ee
where $a^{{\rm cmb}}_{lm}$ are the $a_{lm}$'s of a map generated with CMB signal only and then performing the mask 
projection, and  $a^{{\rm noise}}_{lm}$ are the ones
associated with a map generated with noise only and then performing the mask 
projection. Both of these two matrices have an anisotropic component. For the map 
$\langle a^{{\rm cmb}}_{l_1 m_1} a^{{\rm cmb}}_{l_2 m_2}\rangle$, this arises only because of the 
sky cut, while for the map $\langle a^{{\rm noise}}_{l_1 m_1} a^{{\rm noise}}_{l_2 m_2}\rangle$, 
it is generated both by the sky cut and
by the anisotropy of the noise power. We already discussed about the effect of the anisotropy of the noise in the previous paragraph; this effect turns out to be the most relevant.
The sky cut gives a much smaller effect because, as we will explain more in detail 
in the next Section, it is mainly associated with the average
value of the temperature outside of the sky cut, and this average value is subtracted out before the analysis.

\section{\label{WMAP}Analysis of WMAP 1-year data}

We keep our methodology quite similar to the one used by the WMAP collaboration for their analysis of the 
local shape \cite{Komatsu:2003fd} in order to have a useful consistency check. We compare WMAP data with
Gaussian Monte-Carlo realizations, which are used to estimate the variance of the estimator. 
We generate with HEALPix\footnote{See HEALPix website: http://www.eso.org/science/healpix/} a random 
CMB realization, with fixed cosmological parameters, at 
resolution {\em nside}=256 (786,432 pixels). The
parameters are fixed to the WMAP best fit for a $\Lambda$CDM cosmology with power-law spectrum
\cite{Spergel:2003cb}: $\Omega_b h^2 = 0.024$, $\Omega_m h^2 = 0.14$, $h =0.72$, $\tau = 0.17$, 
$n_s =0.99$. With these parameters,
the present conformal time $\tau_0$ is $13.24$ Gpc and the recombination time is $\tau_R = 0.27$ Gpc. 
A given realization is smoothed with the WMAP window functions for the Q1, Q2, V1, V2, W1, W2, 
W3 and W4 bands \cite{Page:2003eu}. To each of these 8 maps we add an independent noise realization:
for every pixel the noise is a Gaussian random variable with variance $\sigma_0^2/N_{\rm obs}$, where
$N_{\rm obs}$ is the number of observations of the pixel and $\sigma_0$ is a band dependent constant
\cite{Bennett:2003bz}. 
The maps are then combined to give a single map: we make a pixel by pixel average of the 8 maps 
weighted by the noise $\sigma_0^2/N_{\rm obs}$. 
We use this procedure because it is identical to that used in \cite{Komatsu:2003fd}, thereby
allowing direct comparisons between our results and those of the WMAP team.
In future work, it can in principle be improved in two ways.
First of all, for the window function to be strictly rather than approximately uniform across the sky,
the weights of the eight input maps should be constant rather than variable from pixel to pixel.
This is a very small effect in practice, since the eight $N_{\rm obs}$-maps are very similar, making
the weights close to constant.
Second, the sensitivity on small scales can be improved by using $l$-dependent weights \cite{Tegmark:2003}:
for instance, at very high $l$, most of the weight should be given to the $W$ bands, since they have the narrowest beam. 
This would also have a small effect for our particular application, since our estimator uses only the first few hundred multipoles.
We explicitly checked that this would not reduce the variance of our estimator appreciably.
  
We apply the $Kp0$ mask to the final map, to cut out the Galactic plane and the known point sources 
\cite{Bennett:2003ca}: this mask leaves the 76.8\% of the pixels, $f_{\rm sky} = 0.768$. 
The average temperature outside the mask is then subtracted. 

On the resulting map we calculate the estimators defined in the preceding Sections. For the local
shape we performed the analysis both with and without the linear piece discussed in Section \ref{optimal}. The 
quadratic maps of equations (\ref{eq:lin1}) and (\ref{eq:lin2}) used for the linear correction are calculated by
averaging many ($\simeq 600$) Monte-Carlo maps  obtained with the same procedure above. We use
HEALPix to generate and analyze maps at resolution level {\em nside}=256 (786,432 pixels). The integration over $r$ is
performed from $\tau_0-0.025\,\tau_R$ up to $\tau_0-2.5\,\tau_R$ with $\sim 200$ equidistant points, 
and then with another logarithmically spaced $\sim 60$ points up to the present epoch. 
Such a high resolution both in $r$ spacing and in  {\em nside} is necessary in order to reproduce the 
cancellation on squeezed triangles which occurs among the various terms in the equilateral shape (\ref{eq:ours}). The computation
of each $f_{\rm NL}$ on a 2.0 GHz Opteron processor with 2 GB of RAM takes about 60 minutes for the local shape, and 100 minutes for the equilateral shape.

We then apply exactly the same procedure to WMAP data. These maps are analyzed after template foreground
corrections are applied, in order to reduce foreground signal leaking outside the mask, as described in 
\cite{Bennett:2003ca}.

A useful analytic bound on the variance of the estimators for $f_{\rm NL}^{\rm local}$ and $f_{\rm NL}^{\rm equil.}$
is obtained from the variance of the full sky estimator (with homogeneous noise) (\ref{eq:estfull}) 
with an $f_{\rm sky}$-correction which takes into account the reduction in available information from not observing 
the whole sky. 
Taking into account the normalization factor, one readily obtains 
\be
\sigma_{\rm an}^{-2} = f_{\rm sky} \sum_{l_1 < l_2 < l_3} \frac{B_{l_1 l_2 l_3}^2}{C_{l_1} C_{l_2} C_{l_3}} \;,
\label{sigma_an}
\ee    
where $B_{l_1 l_2 l_3}$ must be evaluated for $f_{\rm NL}^{\rm local}$ or $f_{\rm NL}^{\rm equil.}$ equal to 1. 
As we discussed, our approach is not strictly optimal as we are not inverting the full covariance matrix, so we 
expect $\sigma_{\rm an}$ to be smaller than the actual standard deviation that we measure from our Monte-Carlos.

In figure \ref{fig:stdevnl}, we show the standard deviation of estimators for the local shape parameter 
$f_{\rm NL}^{\rm local}$ as a function of the maximum multipole analyzed. 
We compare the results of our Monte-Carlo simulations (with and without the linear correction) with the
analytic bound discussed above. The results without linear correction are compatible with the analysis in \cite{Komatsu:2003fd}. We see that the addition of the linear piece reduces the variance divergence at high $l$'s. The residual divergence is probably associated with the fact that we did not
invert the full covariance matrix. The estimator with the smallest variance is the one with linear 
correction at $l_{\rm max}= 335$, with a standard deviation of $37$. The analytic approximation
has an asymptotic value of $30$ which should be considered the best possible limit with the present data. 
Our estimator thus extracts about $(37/30)^{-2}\approx 66\%$ of the $f_{\rm NL}^{\rm local}$-information (inverse variance) from the WMAP data.
The analysis of the data with 
$l_{\rm max}= 335$ gives $47$, so there is no evidence of deviation from pure Gaussianity and we can 
set a $2\sigma$ limit of
\be
-27 < f_{\rm NL}^{\rm local} < 121  \quad {\rm at} \;95\% \;{\rm C.L.}
\ee

\begin{figure}[ht!]             
\begin{center}
\includegraphics[width=14.0cm]{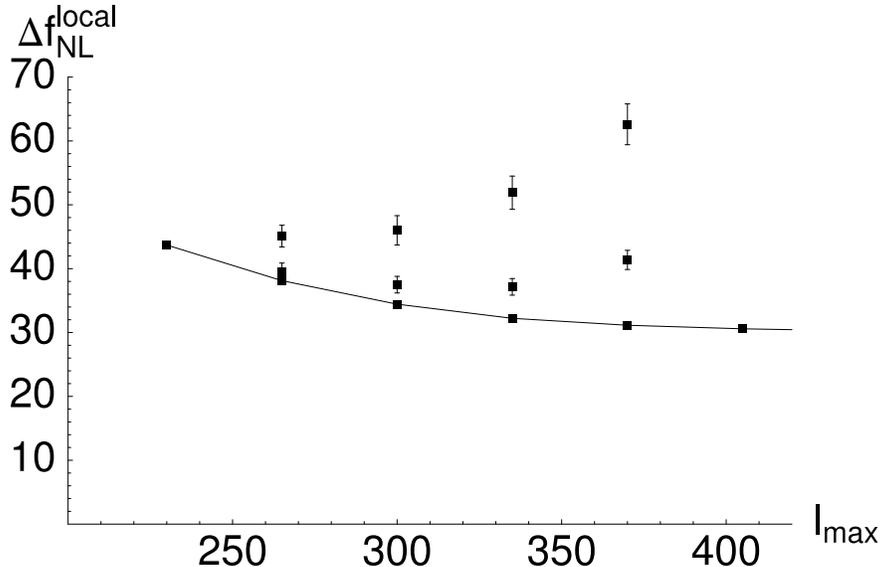}\vspace{0.8cm}
\caption{\label{fig:stdevnl} \small Standard deviation for estimators of $f_{\rm NL}^{\rm local}$ as a 
function of the maximum $l$ used in the analysis. 
Lower curve: lower bound deduced from the full sky variance. Lower data points: standard deviations for the
trilinear + linear estimator. Upper data points: the same for the estimator without linear term, for which the divergence
at high $l$'s caused by noise anisotropy had already been noticed in \cite{Komatsu:2003fd}.}
\end{center}
\end{figure}

In fig.~\ref{fig:noise}, we show a map $S_{AB}(\hat n,r)$ for a radius around recombination 
calculated with noise only and for $l_{{\rm max}}=370$, as an illustrative example of the role of the 
linear piece. As we will clarify below, it is in fact the anisotropy of the noise that causes most
of the contribution to the linear piece. 
The companion map $S_{BB}(\hat n,r)$  is qualitatively very similar to the map  
$S_{AB}$. 
It is the anisotropy of the noise that gives rise to a non-trivial ({\em i.e.}~non-uniform) $S_{AB}$ map, as is clear from the comparison in the same figure 
with a plot of the number of observations per pixel. This can help us in understanding the contribution of the linear
piece associated with the anisotropy of the noise to the estimator. Let us consider a particular Gaussian Monte Carlo realization with a long wavelength mode crossing
one of the regions with a high number of observations. The trilinear piece of the estimator will then detect a spurious
non-Gaussian signal 
associated with the correlation between this long wavelength mode and the small short scale power of the noise 
(\footnote{The effect of this spurious signal obviously averages to zero among many Monte Carlo realizations but it increases the variance of the estimator.}). The maps 
$S_{AB}$ and $S_{BB}$, because of their particular shape,  
will have a non-zero dot product with precisely the same long wavelength mode of the Monte Carlo map, and with an amplitude proportional
to the anisotropy of the two point function of the noise, thus
effectively subtracting the spurious signal detected by the trilinear piece and reducing the variance of the estimator.

\begin{figure}[t!]             
\begin{center}
\includegraphics[width=16.0cm]{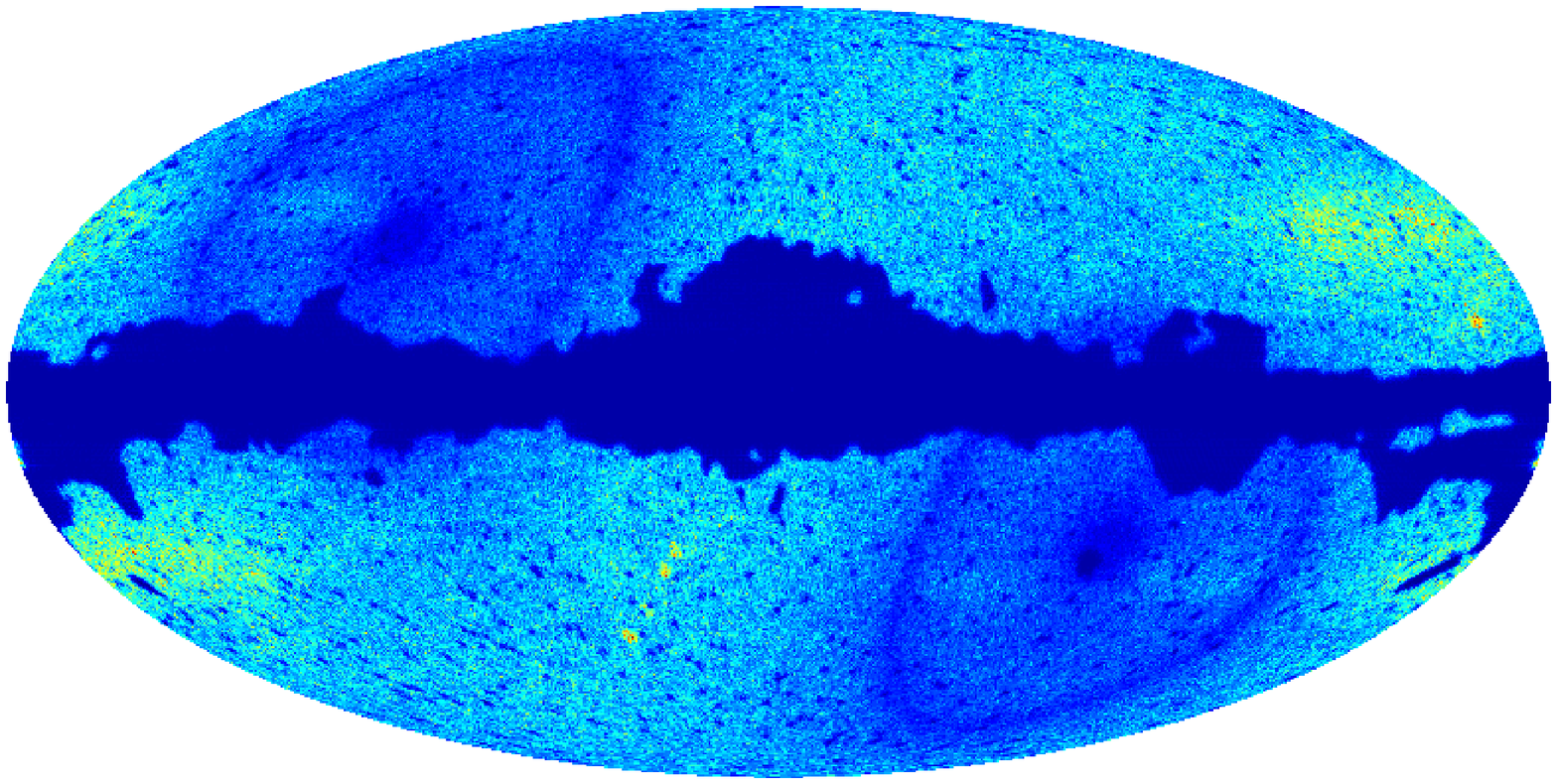}\vspace{0.8cm}
\includegraphics[width=16.0cm]{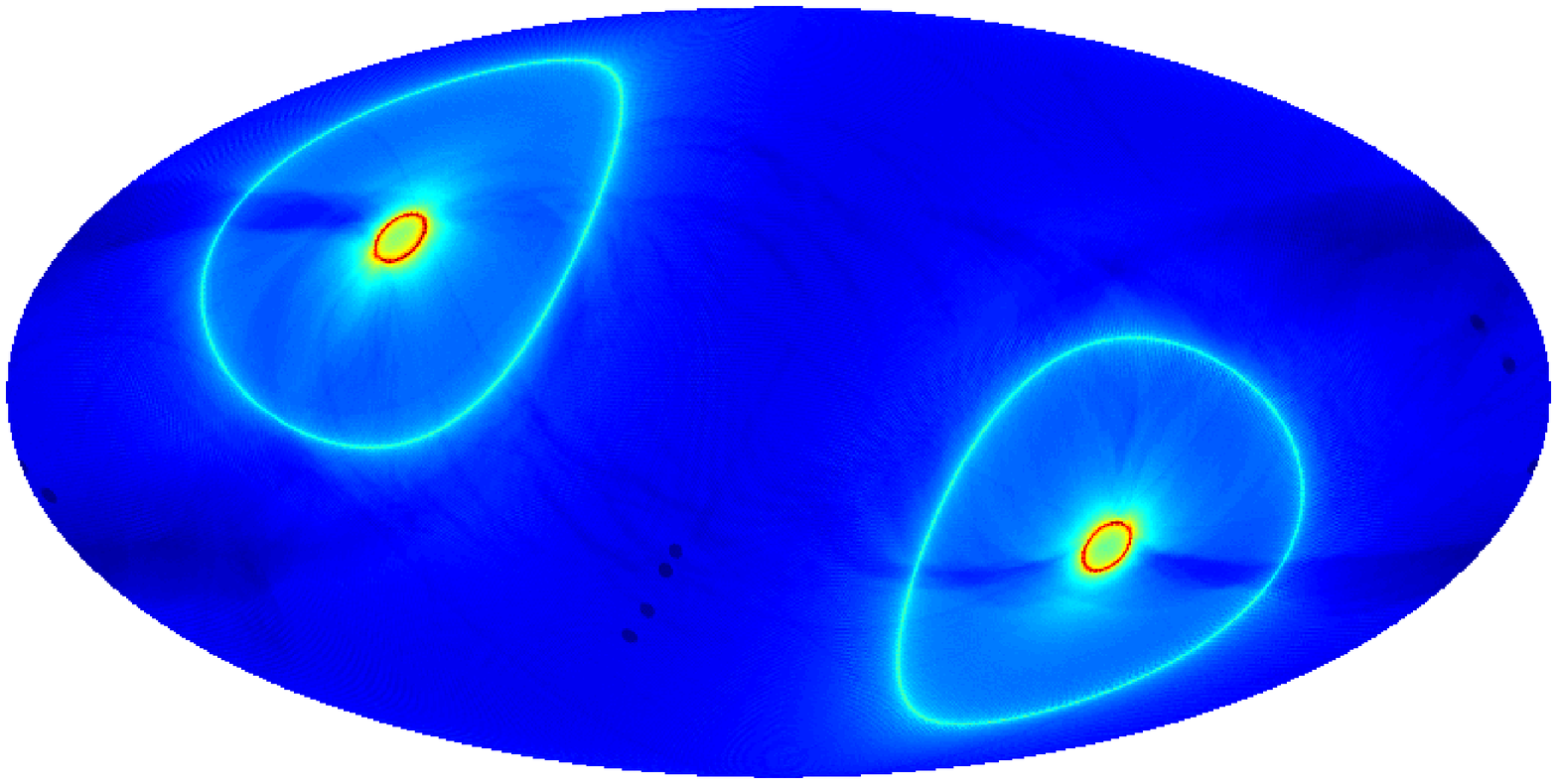}
\caption{\label{fig:noise} \small Top: $S_{AB}(\hat n,r)$ map for $r$ around recombination calculated with noise only.
Bottom: number of observations as a function of the position in the sky for the Q1 band (the plot is quite 
similar for the other bands). A lighter color indicates points which are observed many times.}
\end{center}
\end{figure}

As the mask breaks rotational invariance, the CMB signal also gives non-uniform $S$-maps. However, the breaking 
of rotational invariance can be neglected far from the galaxy mask, so we expect that the contribution to the $S$-maps 
from the CMB is to first approximation constant outside the mask (and zero inside).  In this approximation, the linear
contribution of the estimator would be sensitive only to the average value outside the mask. 
However, given that we are constraining the statistical properties of temperature fluctuations, 
the average temperature in the region outside the mask has to be subtracted before performing the analysis;
therefore we expect the linear correction associated with the CMB signal to be rather 
small. These expectations are in fact verified by our simulations.
We checked that the maps $S$ coming from the CMB signal are to first approximation constant outside the mask, 
with additional features coming from the patches used to mask out bright sources outside the Galactic plane. We then 
verified that if the average value of the temperature is not subtracted, the linear term coming from the CMB signal gives 
a very important reduction of the estimator variance, while its effect is almost negligible when the mean temperature is 
subtracted, as we do in the analysis.

\begin{figure}[t!]             
\begin{center}
\includegraphics[width=14.0cm]{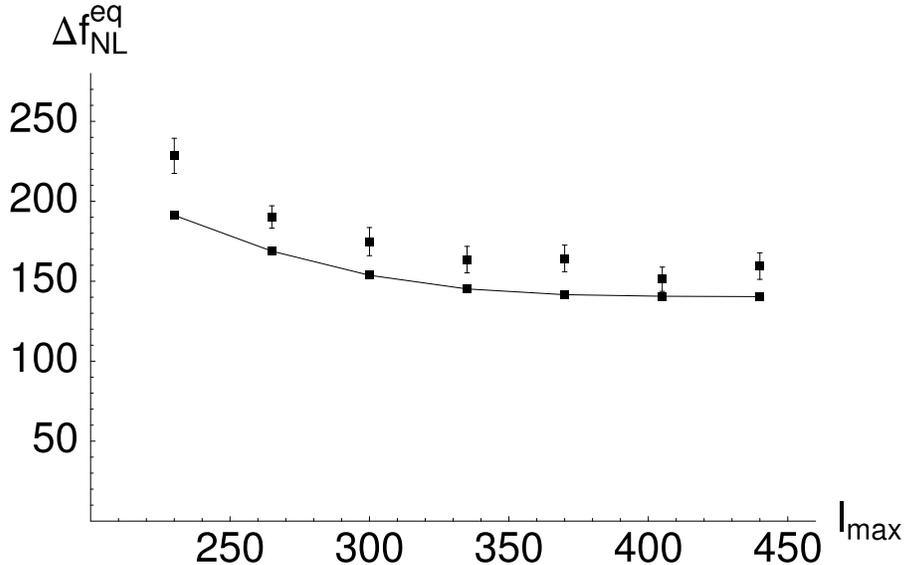}\vspace{0.8cm}
\caption{\label{fig:stdeveq} \small Standard deviations for estimators of $f_{\rm NL}^{\rm equil.}$ as a 
function of the maximum $l$ used in the analysis. 
Lower curve: lower bound deduced from the full sky variance. Data points: standard deviations for the
estimator without linear term.}
\end{center}
\end{figure}

In figure \ref{fig:stdeveq}, we study the standard deviation of the estimators for $f_{\rm NL}^{\rm equil.}$.
The estimator without linear corrections is quite close to the information-theoretical limit, so we did not add the linear corrections (which would require many averages analogous to the maps $S_{AB}$ and $S_{BB}$).
The reason why we have such a good behavior is that, as mentioned, here most of the signal comes from equilateral 
configurations. Thus the estimator is not very sensitive to 
the correlation between the short scale power and the (long wavelength) number of observations.
Moreover the inversion of the covariance matrix 
is important only for the low multipoles, which give only a small fraction of the signal for  $f_{\rm NL}^{\rm equil.}$.  We find that the estimator with smallest standard deviation is at $l_{\rm max}=405$, with a 
standard deviation of $151$. The analysis of the data with the same $l_{\rm max}$ gives $-64$. We deduce a $2\sigma$ limit
\be
-366 < f_{\rm NL}^{\rm equil.} < 238  \quad {\rm at} \;95\% \;{\rm C.L.}
\ee
 The given limit is approximately 2 times stronger than the limit obtained in \cite{Babich:2004gb}, indirectly obtained starting from the limit on $f_{\rm NL}^{\rm local}$. 
The limit appears weaker than for the local case because, for the same $f_{\rm NL}$, the local distribution has 
larger signal to noise ratio than an equilateral one, as evident from fig.~\ref{fig:shapes}. This is not a physical 
difference but merely a consequence of our defining $f_{\rm NL}$ at the equilateral configuration.  

To check that the two limits correspond approximately to the same ``level of non-Gaussianity" we can define a quantity which is sensitive to the 3-point function integrated over all possible shapes of the triangle in momentum space. In this way it will be independent of which point we choose for normalization. We define this quantity, which we will call NG, directly in 3d as an integral of the square of the functions in fig.~\ref{fig:shapes}
\be
\label{eq:NG}
{\rm NG} \equiv \left(\int_\triangle \frac{x_2 d x_2 \, x_3 d x_3}{(2 \pi)^2} \frac{F(1,x_2,x_3)^2}{\Delta_\Phi^3 \, x_2^{-3} x_3^{-3}} \right)^{1/2} \;.
\ee
The integration is restricted to the same triangular region as in fig.s \ref{fig:shapes} and \ref{fig:hddiff}: $1-x_2\leq x_3\leq x_2\le 1$. 
The measure of integration comes from the change of variables from the 3d wavevectors to the ratios $x_2 = k_2/k_1$ and $x_3 =k_3/k_1$. NG is parametrically of order $\Delta_\Phi^{1/2} \cdot f_{\rm NL}$ (where $\Delta_\Phi \simeq 1.9 \cdot 10^{-8}$ \cite{Spergel:2003cb,licia}), which is the correct order of magnitude of the non-Gaussianity, analogous to the skewness for a single random variable. To better understand the meaning of the defined quantity, note that if we integrate inside the parentheses in eq.~(\ref{eq:NG}) over the 
remaining wavevector we get the total signal to noise ratio (or better the non-Gaussian to Gaussian ratio). This further
integration would approximately multiply NG by the square root of the number of data. This means that to detect a given value of NG we need
a number of data of order ${\rm NG}^{-2}$, in order to have a signal to noise ratio of order 1. This is clearly a good way to quantify the deviation of the statistics from pure Gaussianity.   
The $2 \sigma$ windows for $f_{\rm NL}$ can be converted to constraints on NG (we define NG to have the same sign of $f_{\rm NL}$)
\footnote{Note that for the local model the integration in eq.~(\ref{eq:NG}) diverges like the log of the ratio
of the minimum to the maximum scales in the integral. For the quoted numbers we chose a ratio of 1000.}
\be
-0.006 < {\rm NG}^{\rm local}  < 0.025  \quad {\rm at} \;95\% \;{\rm C.L.}
\ee
\be
-0.016 < {\rm NG}^{\rm equil.} < 0.010  \quad {\rm at} \;95\% \;{\rm C.L.}
\ee
Contrary to what one might think from a naive look at the limits on $f_{\rm NL}$,
the maximum tolerated amount of non-Gaussian signal in a map is thus very similar for both shapes.

So far in this paper, we have neglected any possible dependence on the cosmological parameters. A proper analysis would marginalize over our uncertainties in parameters and this would  increase the allowed range of $f_{{\rm NL}}$.
In order to estimate what the effect of these uncertainties is, let us imagine that the real cosmological parameters are not exactly equal to the best fit ones. The cosmological parameters are determined mainly from the 2-point function of the CMB, $C^{{\rm cmb}}_l$, therefore the largest error bars are associated to those combinations of parameters which leave $C^{{\rm cmb}}_l$ unchanged.
In the limit in which the $C^{{\rm cmb}}_l$'s remain the same, also the variance of our estimator (which is computed  with
the best fit cosmological parameters for a $\Lambda$CDM cosmology with power-law spectrum \cite{Spergel:2003cb}) remains
the same: in the weak non-Gaussian limit the variance just depends on the 2-point function.  However, the combination of parameters which leave unchanged the $C^{{\rm cmb}}_l$'s does not necessarily leave unchanged 
the bispectrum, therefore uncertainties in the determination of the cosmological parameters will
have an influence on the expectation value of the estimator. 
The expectation value of the estimator would be:
\be
\langle{{\cal E}}\rangle = \frac{1}{N} \sum_{l_1 < l_2 < l_3} 
\frac{B_{l_1 l_2 l_3} \widetilde B_{l_1 l_2 l_3}}{C_{l_1} C_{l_2} C_{l_3}} \;,
\ee    
where $N$, $B_{l_1 l_2 l_3}$ and $C_{l}$ are the same as in our analysis, computed with
the best fit cosmological parameters, 
while $\widetilde B_{l_1 l_2 l_3}$ is the true bispectrum.
Thus when varying parameters, the normalization $N$ needs to be changed to make the estimator unbiased. 

The most relevant uncertainty is the 
reionization optical depth $\tau$, which is correlated with the uncertainty in the amplitude of the 
power spectrum $\Delta_{\Phi}$. With the purpose of having a rough estimate, we can approximate the effect
of reionization by a multiplicative factor $e^{-\tau}$ in front of the transfer function $\Delta^T_l(k)$
for $l$ corresponding to scales smaller than the horizon at reionization.
In order to keep the $C^{{\rm cmb}}_l$'s unchanged, at least at high $l$'s, we then multiply $\Delta_\Phi$ by
$e^{2\tau}$. The bispectrum is proportional to $\Delta_\Phi^2\ \Delta^T_l(k)^3$ and thus changes even if the $C_l$'s do not. 
For the equilateral shape most of the signal comes from equilateral configurations with all the 3 modes inside the horizon at 
reionization; in this case $\langle{{\cal E}}\rangle$ scales roughly as $e^{\tau}$. For the local shape the signal comes from
squeezed configurations with one mode of much smaller wavelength than the others. Taking only 2 modes inside the horizon at reionization
we get $\langle{{\cal E}}\rangle \propto e^{2 \tau}$.

If we consider the $1\sigma$ error, $\tau=0.166^{+0.076}_{-0.071} $ \cite{Spergel:2003cb}, 
we find that the uncertainty in the reionization depth should
correspond to an error on $f_{\rm NL}$ of order 8\% for the equilateral shape, and 15\% for the local shape. These numbers translate directly into uncertainties in the allowed ranges we quote. 
In the future, if the error induced by the uncertainty in the cosmological parameters will become comparable to the variance of the 
estimator, a more detailed analysis will be required; however, at the moment an analysis with fixed cosmological parameters is certainly 
good enough.

\section{Conclusions}

We have developed a method to constrain the level of non-Gaussianity when the induced 3-point function is of the ``equilateral" type. We showed that the induced shape of the 3-point function can be very well approximated by a factorizable form making the analysis practical. 
Applying our technique to the WMAP first year data we obtained 
\begin{equation}
-366 < f_{\rm NL}^{\rm equil.} < 238  \quad {\rm at} \;95\% \;{\rm C.L.}
\end{equation}
The natural expectation for this amplitude for ghost or DBI inflation  is $f_{\rm NL}^{\rm equil.}
\sim 100$, below the current constraints but at a level that should be attainable in the future. The limit an experiment can set on   $f_{{\rm NL}}$  just scales as the inverse of the maximum $l$ the experiment can detect. 
As a result, in the case of WMAP, increased observing time will approximately decrease the error bars by 30\% and 60\% for 4 years and 8 years of observation. 
The increased angular resolution and smaller noise of Planck pushes the point where noise dominates over signal to $l\sim 1500$. This  should result in a factor of 4 improvement on the present constraints. In addition polarization measurements by Planck can further reduce the range by an additional factor of 1.6  \cite{Babich:2004yc}.

We also constrained the presence of a 3-point function of the ``local" type, predicted for example by the curvaton and variable decay width inflation models, obtaining
\begin{equation}
-27 < f_{\rm NL}^{\rm local} < 121  \quad {\rm at} \;95\% \;{\rm C.L.}
\end{equation}  

We defined a quantity NG, which quantifies for any shape the level of non-Gaussianity of the data, analogously to the 
skewness for a single random variable. The limits on NG are very similar for the two shapes, approximately 
$| {\rm NG} |< 0.02$ at 95\% C.L. . 

We showed that unless one has a full sky map with uniform noise, the estimator must contain a piece that is linear on 
the data in order to extract all the relevant information from the data and saturate the Cramer-Rao bound for the 3-point 
function measurement uncertainty.
This correction is particularly important for the local shape and accounts for the improvement of our limits over that 
from the WMAP team. 
Moreover this correction goes a long way towards reducing the divergence in the variance of the estimator as 
$l_{\rm max}$ is increased.

\section*{Acknowledgments}
We would like to thank Eiichiro Komatsu for his help during the project, 
Eva Silverstein for volunteering to look into foreground contamination if we had found a positive signal, 
and Daniel Babich and Angelica de Oliveira-Costa for help with analysis software.
The numerical analysis necessary for the completion of this paper was performed on the Sauron cluster, 
at the Center for Astrophysics, Harvard University, making extensive use of the HEALPix package \cite{Gorski:1998vw}. 
LS is supported in part by funds provided by the U.S. Department of Energy (D.O.E) under
cooperative research agreement DF-FC02-94ER40818. MZ was supported by the Packard and Sloan foundations, NSF AST-0506556 
and NASA NNG05GG84G.
MT was supported by NASA grant NAG5-11099, NSF grant AST-0134999, the Packard Foundation and Research Corporation.

\end{document}